\documentclass[notitlepage,superscriptaddress]{revtex4}
\usepackage{amsmath,amssymb}
\usepackage{latexsym}
\usepackage[bookmarks=false,colorlinks=true]{hyperref}
\usepackage{bm}
\usepackage{color}

\numberwithin{equation}{section}

\newcommand{\sfrac}[2]{{\textstyle\frac{#1}{#2}}}
\newcommand{\ba}{\begin{array}}
\newcommand{\ea}{\end{array}}
\newcommand{\be}{\begin{equation}}
\newcommand{\ee}{\end{equation}}
\newcommand{\bea}{\begin{eqnarray}}
\newcommand{\eea}{\end{eqnarray}}

\begin{document}
\begin{flushright}
ITP--UH--05/16
\end{flushright}

\title{Spherical Calogero model with oscillator/Coulomb  potential: classical case}

\author{Francisco Correa}
\email{francisco.correa@itp.uni-hannover.de}
\affiliation{Leibniz Universit\"at Hannover, Appelstrasse 2, 30167 Hannover, Germany}
\author{Tigran Hakobyan}
\email{tigran.hakobyan@ysu.am}
\affiliation{Yerevan State University, 1 Alex Manoogian Street, Yerevan, 0025, Armenia}
\affiliation{Tomsk Polytechnic University, Lenin Avenue 30, 634050 Tomsk, Russia}
\author{Olaf Lechtenfeld}
\email{lechtenf@itp.uni-hannover.de}
\affiliation{Leibniz Universit\"at Hannover, Appelstrasse 2, 30167 Hannover, Germany}
\author{Armen Nersessian}
\email{arnerses@ysu.am}
\affiliation{Yerevan State University, 1 Alex Manoogian Street, Yerevan, 0025, Armenia}
\affiliation{Tomsk Polytechnic University, Lenin Avenue 30, 634050 Tomsk, Russia}

\begin{abstract}
We construct the Hamiltonians and symmetry generators of Calogero-oscillator and Calogero-Coulomb models
on the $N$-dimensional sphere within the matrix-model reduction approach.
Our method also produces the integrable Calogero-Coulomb-Stark model on the sphere
and proves the integrability of the spin extensions of all these systems.
\end{abstract}

\maketitle

\section{Introduction}
\noindent
The rational Calogero  model is one of the most important and most known integrable systems invented
in the 20th century~\cite{calogero-review}.
It describes $N$~particles on a line with a pairwise inverse-square interaction potential.
Adding an external oscillator potential preserves its integrability~\cite{calogero0,moser}. Moreover, this model, as well as its  generalizations associated with arbitrary root systems~\cite{calogero-root}, were
 found to be superintegrable (i.e.\ they possess $2N{-}1$ functionally independent constants of motion)~\cite{woj83}.
The hyperbolic version of the Calogero model  remains  superintegrable~\cite{gonera98},
 while its trigonometric counterpart inherits  just the integrability~\cite{sutherland}.
These describe $N$ interacting particles on a circle or hyperbola, correspondingly. Attempts to construct  integrable  analogs of such systems
 in higher dimensions have been unsuccessful see e.g.~\cite{poly97}.
On the other hand, all these models can be interpreted as a single particle in $N$-dimensional Euclidean space
subject to very particular non-isotropic potentials. From this viewpoint it is natural to modify
their $r^{-2}$ radial dependence by putting this particle on an $N$-sphere or -hyperboloid.
This deformation retains the superintegrability but looses the multi-particle interpretation.

In our recent paper~\cite{CalCoul},
we have indicated that the spherical or hyperbolic extensions of the rational Calogero potential
associated with an arbitrary Coxeter group is the only possible superintegrable deformation of the
$N$-dimensional oscillator and Coulomb systems.
We also revealed explicit expressions  for the constants of motion of the Calogero-Coulomb
problem on Euclidean space both in the classical~\cite{CalCoul}  and the quantum~\cite{Runge}  case.
They involve  an analog of the Runge-Lenz vector and its related algebra.
From the other side, the hidden symmetries of the rational Calogero-oscillator
problem have been known for decades and are well investigated \cite{calogero-root,adler}.
The integrable  two-center counterpart of the Calogero-Coulomb system
and an integrable Calogero-Coulomb-Stark system and their symmetry generators
have been constructed  as well~\cite{Calogero-Stark}.

However, it seems that the Calogero-oscillator and Calogero-Coulomb systems extended to the $N$-sphere or -hyperboloid
have not yet been investigated properly.
This is the subject of the current paper.
For arbitrary (positive) Coxeter root systems $\Delta_+=\{\alpha\}$, their Hamiltonians read
 \be
{\cal H} =\frac{\bm{p}^2}{2} \mp \frac{(\bm{x}\cdot\bm{p})^2}{2r^2_0}+
\sum_{\alpha \in \Delta_+}
\frac{g_\alpha^2(\alpha\cdot \alpha )}{2(\alpha\cdot \bm{x})^2} +V(\bm{x})
\label{cc}
\ee
with
\be
V_\omega=\frac{\omega^2r^2_0}{2}\frac{\bm{x}^2}{x_0^2},
\qquad
V_\gamma=-\frac{\gamma}{r_0}\frac{x_0}{|\bm{x}|}
\qquad\textrm{where}\qquad
x^2_0 \pm \bm{x}^2 = r^2_0.
\label{potentials}
\ee
Here, the upper/lower sign corresponds to the sphere/hyperboloid, $\bm{x}$ and $x_0$ are Cartesian coordinates
in the ambient $(N{+}1)$-dimensional space, and $r_0$ is the radius of the $N$-dimensional sphere/hyperboloid.
The vectors $\alpha$ from the set $\Delta_+$ of positive roots uniquely characterize the Coxeter reflections,
and the coupling constants $g_\alpha$ form a reflection-invariant discrete function.
The original Calogero potential corresponds to the $A_{N-1}$ Coxeter system with
the positive roots given in terms of the standard basis by $\alpha_{ij}=e_i-e_j$
for $i<j$. The reflections become the coordinate
permutations in this particular case.

In the absence of the Calogero interaction ($g_\alpha=0$),
such systems are reduced to the spherical and pseudo-spherical Coulomb and oscillator  systems, introduced a long time ago
by Schr\"odinger and Higgs~\cite{schroedinger,higgs}. These systems have  direct analogs of the
hidden-symmetry generators  of their Euclidean counterparts. However, their symmetry algebras are essentially
nonlinear even when restricted to a constant energy surface. In the absence of  the Calogero and
oscillator/Coulomb potentials, the above Hamiltonian describes a free particle on the $N$-dimensional sphere/hyperboloid.
Hence, the system \eqref{cc} with vanishing potential $V(\bm{x})$ is nothing but the spherical/hyperbolic
Calogero model associated  with an arbitrary reflection group.
Particle motion near the horizon of an  extremal $(2N{+}1)$-dimensional Perry-Myers black hole  is an
example of such a system~\cite{gns}.

We remark that the so-defined spherical Calogero model differs from the
 angular Calogero model investigated recently in~\cite{sphCal,sph-mat,flp,Feigin}.
More precisely,  the conventional  Calogero model on the $N$-dimensional sphere is given by  the Hamiltonian
\be
\label{C0}
{\cal H}_0 =\frac{\bm{p}^2}{2}- \frac{(\bm{x}\cdot\bm{p})^2}{2r^2_0}+
\sum_{i<j}
\frac{g^2}{(x_i-x_j)^2},
\ee
while the angular Calogero Hamiltonian~${\cal H}_\Omega$ corresponding to the angular part of the Calogero system
on $(N{+}1)$-dimensional flat space is defined by
\be
r_0^{-2} {\cal H}_\Omega = \frac{\bm{p}^2}{2}- \frac{(\bm{x}\cdot\bm{p})^2}{2r^2_0}+
\sum_{i<j}
\frac{g^2}{(x_i-x_j)^2} +\sum_{i=1}^N\frac{g^2}{(x_i-x_0)^2}
\qquad\textrm{with}\qquad x_0=\sqrt{r_0^2-\bm{x}^2}.
\label{angular}
\ee

The goal of the current paper is to  investigate the symmetries of the  Calogero-oscillator and
Calogero-Coulomb  systems on the $N$-dimensional sphere using the matrix-model reduction.
For this purpose first we reformulate  the $N^2$-dimensional spherical oscillator/Coulomb system
on the space of Hermitian $N\times N$ matrices. Then, reducing this system by the SU$(N)$ adjoint action
and fixing  specific values of its  generators in a standard way, we  arrive at  the desired
Calogero-oscillator and Calogero-Coulomb systems \eqref{cc}.
The SU$(N)$ invariant polynomials in the matrix-model symmetry generators yield
the correct integrals of motion  for the reduced  system.
In a similar way we find an integrable Calogero-Coulomb-Stark system on the sphere.
We  restrict ourselves to the spherical $A_{N-1}$ Calogero model~\eqref{C0} supplemented by
the potentials~\eqref{potentials}.
We do not consider the hyperboloid case since the transition from the spherical one is straightforward.

The paper is organized as follows.
In Section~2 we present general properties of the classical oscillator and Coulomb systems on
the sphere in a parametrization relevant for our purposes.
In Section~3 we construct the symmetry generators of  the classical Calogero-oscillator and
Calogero-Coulomb systems on the $N$-dimensional sphere using the matrix-model reduction.
In Section~4 we present the Calogero-Coulomb-Stark system  and briefly discuss the
spin generalizations of considered models.

\section{Preliminary: oscillator and Coulomb systems on the sphere}
\noindent
Let us briefly describe,  following~\cite{higgs}, the Coulomb and oscillator systems
on an $N$-dimensional sphere of radius~$r_0$, parameterized $N$~Cartesian coordinates
of the ambient $(N{+}1)$-dimensional space.
In this parametrization, the sphere metric can immediately be obtained by the restriction
of the flat metric on $\mathbb{R}^{N+1}$,
\be
ds^2=h_{ij}dx_idx_j=d{\bm{x}}^2 +dx_{0}^2\vert_{x^2_0 + \bm{x}^2 = r^2_0}
=d\bm{x}^2+\frac{(\bm{x}\cdot d\bm{ x})^2}{r^2_0-\bm{ x}^2}.
\label{metric}\ee
The phase space of the  systems on the sphere is given by its cotangent bundle equipped with
the canonical symplectic structure $d\bm{ p}\wedge d\bm{ x}$ and the canonical Poisson brackets
$\{p_i,x_j\}=\delta_{ij}$.
In these terms, the SO$(N{+}1)$ isometries of the sphere are given by the  generators
$(L_{\mu\nu})=(L_{0i},L_{ij})$ via
\be
\begin{gathered}
L_{0i}=x_0 p_i, \qquad
L_{ij}=x_ip_j-x_jp_i
\qquad\textrm{and}\qquad
\{L_{\mu\nu},L_{\rho\lambda}\}=\delta_{\mu\lambda}L_{\nu\rho}+\delta_{\nu\rho}L_{\mu\lambda}
-\delta_{\mu\rho}L_{\nu\lambda}-\delta_{\nu\lambda}L_{\mu\rho},
\label{Lij}
\end{gathered}
\ee
where $x_0=\sqrt{r_{0}^2-\bm{x}^2}$.
The above  $so(N{+}1)$ algebra decomposes as
\begin{gather}
\label{poisL0j}
\{L_{0i},L_{0j}\}=L_{ij},
\qquad \{L_{0i},L_{kj}\}=\delta_{ik}L_{0j}-\delta_{ij}L_{0k},
\\
\label{poisLij}
\{L_{ij},L_{kl}\}=
\delta_{il}L_{jk}+\delta_{jk}L_{il}-\delta_{ik}L_{jl}-\delta_{jl}L_{ik},
\end{gather}
where the $L_{ij}$ generate the $so(N)$ subalgebra.

\bigskip

\emph{The oscillator  on the $N$-dimensional sphere} is defined by the Hamiltonian~\cite{higgs}
\be
\label{Hx}
H_\omega
=\frac{\bm{p}^2}{2}-\frac{(\bm{x}\cdot\bm{p})^2}{2r_{0}^2} + \frac{\omega^2 r_{0}^2  }{2} \frac{\bm{x}^2}{x_0^2}.
\ee
The  symmetries  of this Hamiltonian are given by the generators of
the SO$(N)$ angular momentum $L_{ij}$  defined in~\eqref{Lij} and the
hidden-symmetry generators
\be
\label{Iij}
I_{ij}=
\frac{x_0^2}{r_{0}^2} p_ip_j+ \frac{\omega^2r_{0}^2}{x_0^2}x_jx_j.
\ee
Note that  these expressions can be obtained from those of the flat oscillator  by the replacement
  \be
  x_i\to \frac{r_0}{x_0} x_i,
  \qquad
  p_i\to \frac{L_{0i}}{r_0}=\frac{x_0}{r_0}p_i.
  \label{rep}
  \ee
The  symmetry
algebra of the spherical system is essentially nonlinear \cite{higgs},
\begin{gather}
\label{poisLI}
\{L_{ij},I_{kl}\}=\delta_{jk}I_{il}-\delta_{il}I_{jk}-\delta_{ik}I_{jl}+\delta_{jl}I_{ik},
\\
\label{poisII}
\{I_{ij},I_{kl}\}=
-\omega^2(\delta_{il}L_{jk}+\delta_{jk}L_{il}+\delta_{ik}L_{jl}+\delta_{jl}L_{ik})
-\frac{1}{r_{0}^2}(I_{il}L_{jk}+I_{jk}L_{il}+I_{ik}L_{jl}+I_{jl}L_{ik}).
\end{gather}
The Hamiltonian can be expressed in terms of the symmetry generators,
\be
H_\omega =\frac12\sum_i I_{ii}+\frac{\bm{L}^2}{2r_{0}^2},
\qquad{\rm where}\qquad
\bm{L}^2=\sum_{i<j}L_{ij}^2.
\ee

In the $r_0\to \infty$ limit the generators $L_{ij}$ and $I_{ij}$ reduce, respectively, to the angular momentum
and Fradkin tensors. Together they form
the SU$(N)$ Poisson bracket algebra, which describes the symmetries of the
standard $N$-dimensional oscillator \cite{fradkin}.

\bigskip

\emph{The Coulomb system on the $N$-dimensional sphere} possesses a similar structure~\cite{higgs}.
It is defined by the Hamiltonian
\be
\label{Hr}
H_\gamma =\frac{\bm{p}^2}{2}-\frac{(\bm{x}\cdot\bm{p})^2}{2r_{0}^2} - \frac{\gamma}{r_0}\frac{x_0}{x}
\qquad\textrm{with}\qquad x=\sqrt{\bm{x}^2}.
\ee
Its  symmetry involves the  SO$(N)$ angular momentum  tensor $L_{ij}$  and the analog
of the Runge-Lenz vector,
\be
\label{Ai}
A_i = \frac{x_0}{r_0}\sum_j  L_{ij} p_j -\frac{\gamma x_i}{x}.
\ee
The square of the latter involves the energy,
\be
\label{Asq}
\bm{A}^2=\left(2H_\gamma -\frac{\bm{L}^2}{r_{0}^2}\right) \bm{L}^2+\gamma^2.
\ee
It is easy to see  that the expression for the Runge-Lenz vector on the sphere
can again be obtained from the one on flat space by the replacement \eqref{rep}.
The symmetry algebra  is  given by the relations \eqref{poisLij}
and by
\be
\label{poisLA}
\{L_{ij},A_k\}=-\delta_{ik}A_j+\delta_{jk}A_i,\qquad
\{A_i,A_j\}=2\left(H_\gamma-\frac{\bm{L}^2}{r_{0}^2}\right) L_{ij}.
\ee

In the flat-space limit and on a fixed-energy level it is reduced to the SO$(N,1)$ or SO$(N{+}1)$ symmetry
for positive or negative energy, correspondingly.

\section{Spherical Calogero-Coulomb and Calogero-oscillator from matrix models}
\noindent
In this section we construct the spherical Calogero-oscillator and Calogero-Coulomb  systems
using the matrix-model reduction.
More precisely, we consider the usual  spherical oscillator and Coulomb systems
on the space of $N\times N$ Hermitian matrices
and then reduce them by the adjoint SU$(N)$ action.
As a result, we get the spherical Calogero-oscillator and Calogero-Coulomb
systems suggested in~\cite{CalCoul}. This approach allows us
to find explicit expressions for all symmetry generators (including the  hidden ones).
In addition, it is immediately generalized to
superintegrable spin extensions of these models, as well as to an integrable
spherical Calogero-Coulomb-Stark model.

\subsection{Calogero-oscillator on the sphere}
\noindent
Let us define the Hermitian matrix model for the oscillator on the sphere by the Hamiltonian
\be
\label{Hx-mat}
{\cal H}_{\omega}^\text{mat}  =
\frac{1}{2} \text{tr}\,\mathbf{P}^2 - \frac{1}{2r_{0}^2} (\text{tr}\,\mathbf{PX})^2
+ \frac{\omega^2 r_{0}^2}{2}\frac{\text{tr}\,\mathbf{X}^2}{r_{0}^2-\text{tr}\, \mathbf{X}^2 }.
\ee
Here $\mathbf{P}$ and $\mathbf{X}$ are
Hermitian matrices containing, respectively, $N^2$ momenta and coordinates:
\be
\label{PX}
\mathbf{P} = \sum_{a=1}^{N^2} P_a \mathbf{T}_a,
\qquad
\mathbf{X} = \sum_{a=1}^{N^2} X_a \mathbf{T}_a.
\ee
We have introduced a basis of orthonormalized U$(N)$ generators,
\be
\label{Ta}
[\mathbf{T}_a,\mathbf{T}_b]=\imath\sum_a f_{abc}\mathbf{T}_c,
\qquad
\text{tr}\, \mathbf{T}_a\mathbf{T}_b=\delta_{ab}.
\ee
Their  explicit form  can be set, in particular, by choosing $N$ matrices to be diagonal,
\be
\label{Ti}
\mathbf{T}_{(i-1)N+i}=\mathbf{E}_{i,i},
\qquad 1\le i\le N,
\ee
where $\mathbf{E}_{i,j}$ has vanishing entries except for one in the $i$-th row and $j$-th column:
\be
[\mathbf{E}_{i,j}]_{i'j'}=\delta_{ii'}\delta_{jj'}.
\ee
The remaining $\mathbf{T}_a$ are  selected from the following set of the $N(N-1)$ off-diagonal matrices:
\be
\label{Tij}
\mathbf{T}_{(j-1)N+i}=\frac{1}{\sqrt{2}}(\mathbf{E}_{j,i} + \mathbf{E}_{i,j} ),
\qquad
\mathbf{T}_{(i-1)N+j}=\frac{\imath}{\sqrt{2}}(\mathbf{E}_{j,i} - \mathbf{E}_{i,j} ),
\qquad
1\le j<i\le N.
\ee

In terms of the phase-space variables $(P_a,X_a)$, the equivalence of the  matrix model \eqref{Hx-mat}
to the Hamiltonian of the $N^2$-dimensional spherical oscillator \eqref{Hx} becomes transparent:
\be
\label{HxPX}
H_\omega^\text{mat}
=\frac12 \sum_{a}P^2_a-\frac{1}{2r^2_0}\left(\sum_{a}X_aP_a\right)^2+ \frac{r_0^2}{x^2_0}\frac{\omega^2x^2}{2}
\ee
%}
with $x$ and $x_0$ defined by
\be
\label{x-x0}
x^2=\sum_{a=1}^{N^2} X_a^2,\qquad x_0=\sqrt{r^2_0-x^2}.
\ee
According to the previous section, the above system remains invariant under the action
of  the angular momentum and hidden symmetry generators
\be
\label{Lab}
L_{ab}=X_aP_b-X_bP_a,
\qquad
I_{ab}=\frac{x_0^2}{r_{0}^2} P_aP_b+ \frac{\omega^2r_{0}^2}{x_0^2} X_aX_b.
\ee
These constants of motion obey Poisson bracket relations similar to~\eqref{poisLij}, \eqref{poisLI} and \eqref{poisII}.
They can be presented in matrix form,
\begin{align}
\label{L}
\mathbf{L}&=\sum_{a,b}L_{ab}\mathbf{T}_a\otimes\mathbf{T}_b
    =\mathbf{X}\wedge\mathbf{P},
\\
\label{I}
\mathbf{I}&=\sum_{a,b} I_{ab}\mathbf{T}_a\otimes \mathbf{T}_b
    =  \frac{x^2_0}{r^2_0}\mathbf{P}\otimes\mathbf{P} +\frac{\omega^2r_{0}^2}{x_0^2} \mathbf{X}\otimes\mathbf{X}.
\end{align}
The matrix Hamiltonian \eqref{Hx-mat} remains invariant under the adjoint SU$(N)$ action
\be
\label{SUn}
\mathbf{P}\to\mathbf{ U}\mathbf{P}\mathbf{U}^+,
\qquad
\mathbf{X}\to \mathbf{U}\mathbf{X}\mathbf{U}^+.
\ee
The related Noether current is given by a traceless Hermitian matrix with $su(N)$-valued entries,
\be
\label{J}
\mathbf{J}=\imath[\mathbf{X},\mathbf{P}],\qquad J_a=\text{tr}\,\mathbf{J}\mathbf{T}_a=-\sfrac12\sum_{b,c} f_{abc}L_{bc},
\qquad
\{J_a,J_b\}=\sum_c f_{abc} J_c.
\ee
To perform the reduction, we  diagonalize
the coordinate  matrix  by the use of an SU$(N)$ transformation~\eqref{SUn}. Then we  fix the level surface
$\mathbf{J}={\sf const}$ and take into account that the diagonalization of the coordinate matrix
leads to the vanishing of the diagonal entries of $\mathbf{J}$.
The following level surface reproduces   the Calogero potential,
\be
J_{ij}=g(\delta_{ij}-1).
\label{Jijcal}
\ee
As a result, the phase-space variables are mapped to \cite{kazhdan,calogero-root,calogero-review}
\be
\label{XPred}
X_{ij}= x_i\delta_{ij}
\qquad
\textrm{and}
\qquad
P_{ij}=
p_i\delta_{ij}+\imath g\frac{1-\delta_{ij}}{x_i-x_j}\, .
\ee
Their diagonal entries  define the coordinates and
momenta of the $N$-dimensional Calogero system.
Only these preserve the canonical Poisson bracket  relations. Nevertheless, we keep
the same notation for the reduced matrices  \eqref{XPred}.
Note that the reduced momentum matrix $P_{ij}$ corresponds to the Lax matrix
of the standard Calogero system \cite{moser}.

Using the expressions for  the U$(N)$ generators \eqref{Ti} and \eqref{Tij},
we get an explicit form of the orthogonal coordinates~\eqref{PX}  under the reduction,
\be
\label{XaPa}
X_{(j-1)N+i}=
\begin{cases}
x_i & \text{for $i=j$,}
\\
0 & \text{for $i\ne j$,}
\end{cases}
\qquad\qquad
P_{(j-1)N+i} =
\begin{cases}
p_i & \text{for $i=j$, }
\\
\displaystyle
\frac{\sqrt{2}g}{x_i-x_j} & \text{for $i<j$, }
\\
0  & \text{for $i>j$. }
\end{cases}
\ee
Consequently, the matrix model \eqref{Hx-mat} is reduced  to the
Calogero-oscillator model  on the $N$-dimensional sphere,
\be
\label{Hx-red}
\mathcal{H}_\omega=\frac{\bm{p}^2}{2}-\frac{(\bm{x}\cdot\bm{p})^2}{2r_{0}^2}
+\sum_{i<j}\frac{g^2}{(x_i-x_j)^2}+\frac{\omega^2r_{0}^2}{2}\frac{x^2}{x_0^2}.
\ee

Thus, the above system can be obtained from the $N^2$-dimensional Higgs oscillator by
projecting into the orbits of the adjoint group action \eqref{SUn}.
This projection breaks the Poisson structure:
it respects the brackets  only among the observables invariant with respect to the reduction group.
Therefore,  only the SU$(N)$ invariant constants of motion of the matrix system remain constants
in the projected system.
They are built  by taking  an appropriate combination of the elements \eqref{Lab}, which are preserved
under the group action \eqref{SUn}.
In particular,  one can construct the following two sets of SU$(N)$ invariants from the angular momentum
and hidden-symmetry generators:
\be
\label{LkIk}
\mathcal{L}_{2k}=(\text{tr}\otimes\text{tr})\,\mathbf{L}^{2k},
\qquad
\mathcal{I}_k=(\text{tr}\otimes \text{tr})\, \mathbf{I}^k,
\qquad
k=1,2,3,\ldots,
\ee
where the left (right) trace from the tensor product $\text{tr}\otimes \text{tr}$ is performed over the left (right)
factor in the tensor products \eqref{L} and \eqref{I}.
Substituting the explicit  expressions obtained from \eqref{L}, \eqref{I}, and \eqref{XPred}
into the above equations, we arrive at the integrals of motion of the Calogero-oscillator
system on sphere \eqref{Hx-red} . Note that the integrals  $\mathcal{L}_{2k+1}$ vanish
since the angular momentum tensor is antisymmetric.

Already the above integrals form an overcomplete set of constants of motion,
with enough functionally-independent ones to
ensure the superintegrability of the system~\eqref{Hx-red} discovered in~\cite{CalCoul}.
Nevertheless, we can construct constants  of motion  from  more general SU$(N)$ invariants.
They are obtained from $L_{ab}$ and $I_{ab}$ using the SU$(N)$ invariant tensors in the adjoint representation,
\be
d_{a_1\dots a_l}=\text{tr}\, \mathbf{T}_{a_1}\dots \mathbf{T}_{a_l}.
\label{inv}
\ee
According to \eqref{Ta}, \eqref{Ti} and \eqref{Tij},
\be
\label{inv12}
d_a=\sum_{k=1}^N\delta_{ka},
\qquad
d_{ab}=\delta_{ab},
\qquad
d_{abc}-d_{bac}=\imath f_{abc}.
\ee
Note that the vector $d_a$ projects to the U$(1)$ generator of the unitary group
given by the identity matrix.
The higher-rank tensors are more complicated.
As a consequence of the  completeness  condition
obeyed by the basis~\eqref{Ta}, they may be expressed in terms of the $d_{abc}$ by a successive application
of the  formula
\be
\sum_c d_{a_1\dots a_k c}d_{cb_1\dots b_l}	= d_{a_1\dots a_k b_1\dots b_l }.
\ee
Here is an example of  a mixed fourth-order (in momenta) invariant,
\be
\sum_{a,a',\dots}d_{abc}d_{a'b'c'}L_{aa'}L_{bb'}I_{cc'}.
\ee
In this context, the integrals \eqref{LkIk} can be defined alternatively as
\begin{align}
\mathcal{L}_{2k}
&=\sum_{a_1,b_1\dots,a_{2k},b_{2k}} d_{a_1\dots a_{2k}}d_{b_1\dots b_{2k}}L_{a_1b_1}\ldots L_{a_{2k}b_{2k}},
\label{Lk}
\\
\mathcal{I}_k
&=\sum_{a_1,b_1\dots,a_k,b_k} d_{a_1\dots a_k}d_{b_1\dots b_k}I_{a_1b_1}\ldots I_{a_kb_k}.
\label{Ik}
\end{align}
By contracting adjacent indices with a Kronecker delta,
we arrive at  another simple set of invariants,
\begin{align}
\label{Lk'}
{\cal L}'_{2k}&=\sum_{a_1\dots a_k} L_{a_1a_2}L_{a_2a_3}\dots L_{a_{2k}a_1}
=\text{Tr} \;{\rm L}^{2k},
\\
\label{Ik'}
{\cal I}'_k&=\sum_{a_1\dots a_k} I_{a_1a_2}I_{a_2a_3}\dots I_{a_ka_1}
=\text{Tr} \;{\rm I}^k.
\end{align}
Here the rank-four tensors \eqref{L} and \eqref{I} are treated as  ordinary  matrices with entries $I_{ab}$
 and $L_{ab}$ correspondingly, i.e.\  $[{\rm I}]_{ab}=I_{ab}$ and $[{\rm L}]_{ab}=L_{ab}$. Their counterparts
with ``free boundaries'' are obtained by saturating the first and last indices with the invariant vector $d_a$.
For example,
\be
\label{Ik''}
{\cal I}''_k=\sum_{i,a_2\dots a_k,j} I_{ia_2}I_{a_2a_3}\dots I_{a_kj}
=\sum_{i,j=1}^N\bigl({\rm I}^k\bigr)_{ij}.
\ee

Let us derive now the explicit form for the second-order (in momentum) integrals.
The  invariant $\mathcal{L}_2$ corresponds to the angular part  of the pure Calogero model,
whose integrals of motion have been constructed already~\cite{sph-mat}.
It can be presented in terms of the momentum and angular momentum \eqref{Lij} as
\be
\mathcal{L}_2= 2\sum_{i<j} \left( L_{ij}^2 +\frac{2g^2 x^2}{(x_i-x_j)^2}\right).
\ee
It is equivalent to the $(N{-}1)$-dimensional Hamiltonian $H_\Omega$ \eqref{angular}
from the Introduction with the replacement $x\to r_0$ and $x_N\to x_0$.

Of course, the constructed integrals are subject to algebraic relations
since only $2N{-}1$  of them are functionally independent.
The first members of the families \eqref{Ik} and \eqref{Ik''} coincide and are equal to
\be
\label{calXP}
\mathcal{I}_1=
\mathcal{I}''_1 = \left(\frac{x_0{\cal P}}{r_0}\right)^2+\left(\frac{\omega r_0{\cal X}}{x_0}\right)^2
\qquad
\text{with}
\qquad
{\cal X}=\sum_{i=1}^N x_i,
\qquad
{\cal P}=\sum_{i=1}^N p_i.
\ee
It is easy to see also that ${\cal L}'_2=-{\cal L}_2$ and ${\cal I}'_2={\cal I}_2$.
The   Hamiltonian itself is expressed through the constructed integrals as
\be
\mathcal{H}_\omega=\frac12\mathcal{I}'_1+\frac{\mathcal{L}_2}{4r_0^2}.
\ee
Moreover, the families \eqref{Lk'}, \eqref{Ik'} and \eqref{Ik''} do not give rise to
new constants of motion. Indeed, the rank-two matrices $L_{ab}$ and $I_{ab}$
are subject to the third-order relations
\begin{align}
{\rm L}^3+\frac{{\cal L}_2}{2} {\rm L}=0,
\qquad
{\rm I}^3 -{\cal I}'_1 {\rm I}^2+\frac{\omega^2{\cal L}_2}{2} {\rm I}=0,
\end{align}
and the surviving first members are already given by~\eqref{calXP}.

In the flat-space limit, $r_0\to \infty$, we obtain the Calogero model with oscillator
potential \cite{calogero0}.
The generators \eqref{L} and \eqref{I} constitute the pure U$(N^2)$ symmetry
of the initial matrix Hamiltonian ${\cal H}^\text{mat}_\omega$.
Apart from the Noether integrals \eqref{J}, the other SU$(N)$ integrals may be expressed
in terms of matrix analogs of holomorphic and antiholomorphic variables \cite{calogero-review}.
In terms of the current phase variables, these quantities  are
\be
\mathbf{C}=\frac1{2\omega}\mathbf{P}^2+\frac\omega2\mathbf{X}^2+ \frac12\mathbf{J},
\qquad
C_c =\frac{\omega}{2} J_c+\frac1{2\omega}\sum_{a,b} d_{abc} I_{ab},
\qquad
\{C_a,C_b\}=\sum_c f_{abc} C_c.
\ee
According to our procedure,  the constants of motion of the reduced Hamiltonian ${\cal H}_\omega$
for $r_0\to\infty$, corresponding to the aforementioned integrals, can be constructed in the following way:
\be
\label{Cn}
\mathcal{C}_k=\text{tr}\,\mathbf{C}^k
= \sum_{a_1,\dots,a_k} d_{a_1\dots a_k}C_{a_1}\ldots C_{a_k}.
\ee
In fact, $\mathcal{C}_k$ is a $k$-th order Casimir element of the $su(N)$
Poisson algebra  for $1\le k\le N$.
 Therefore, they are in involution, $\{{\cal C}_k,{\cal C}_l\}=0$, and constitute a system of Liouville  integrals
for the Calogero model in the external oscillator potential \cite{calogero-review}.
The first element coincides with the
Hamiltonian:  ${\cal H}_\omega=\omega{\cal C}_1$.

\subsection{Calogero-Coulomb on the sphere}
\noindent
The Hermitian matrix model  for the  Coulomb system on the sphere is defined by the Hamiltonian
\be
\label{Hr-mat}
\mathcal{H}_\gamma^\text{mat}  =
\frac{1}{2} \text{tr}\,\mathbf{P}^2 - \frac{1}{2r_{0}^2} (\text{tr}\,\mathbf{PX})^2
- \gamma\left(\frac{1}{\text{tr}\mathbf{X}^2 }-\frac{1}{r_0^2}\right)^\frac12.
\ee
%}
In terms of the phase-space variables $(P_a,X_a)$, the matrix Hamiltonian~\eqref{Hr-mat}
becomes identical to the Hamiltonian of the spherical Coulomb system~\eqref{Hr},
\be
\label{HrPX}
H_\gamma^\text{mat}=\frac12 \sum_{a}P^2_a-\frac{1}{2r^2_0}\left(\sum_{a}X_aP_a\right)^2 - \frac{\gamma}{r_0}\frac{x_0}{x}
\ee
with $x$ and $x_0$ defined by \eqref{x-x0}.
Apart from the kinematical  angular momentum tensor \eqref{L}, it has a conserved  Runge-Lenz vector
\eqref{Ai}  given by
\be
\label{Aa}
A_a = \frac{x_0}{r_0}\smash{\sum_b} L_{ab}P_b - \frac{\gamma X_a}{x}.
\ee
The latter can be presented through the Hamiltonian,
\be
\label{Aa-2}
A_a=\left(\frac{2x_0}{r_0}\mathcal{H}^\text{mat}_\gamma+\frac{x_0}{r_0^3}(\bm{x}\cdot\bm{p})^2-\frac{2\gamma x}{r_0^2}+\frac{\gamma}{x}\right) X_a
-\frac{x_0}{r_0}(\bm{x}\cdot\bm{p}) P_a,
\ee
and the corresponding matrix reads
\be
\label{Amat}
\mathbf{A} = \sum_aA_a\mathbf{T}_a=\frac{x_0}{r_0}\left( \mathbf{X}\, \text{tr}\,\mathbf{P}^2
- \mathbf{P}\ \text{tr}\, \mathbf{XP}\right) -\frac{\gamma}{x} \mathbf{X}.
\ee
The symmetry generators $L_{ab}$ and $A_a$ obey the Poisson brackets of the Coulomb
system on the sphere, see~\eqref{poisLij} and~\eqref{poisLA}.

The SU$(N)$ reduction procedure implemented above for the Calogero-oscillator system on the sphere
remains valid in this case too.
The matrix-model Hamiltonian \eqref{Hr-mat} can be reduced by the symmetry
\eqref{SUn} to the Calogero-Coulomb  model  on the $N$-dimensional sphere,
\be
\label{HrCal}
\mathcal{H}_\gamma
=\frac{\bm{p}^2}{2}-\frac{(\bm{x}\cdot\bm{p})^2}{2r_0^2}
+\sum_{i<j}\frac{g^2}{(x_i-x_j)^2}- \frac{\gamma}{r_0}\frac{x_0}{x},
\ee
where the matrix coordinates $P_a$ and $X_a$ are defined in~\eqref{XaPa}.

The symmetries of the reduced system are described by symmetric polynomials in the reduced angular momentum
$L_{ab}$ and Runge-Lenz vector $A_c$, with indices coupled by the SU$(N)$ invariant tensors~\eqref{inv}
as in the example below,
\be
\sum_{a,a',\dots}d_{abc}d_{a'b'c'}L_{aa'}L_{bb'}A_cA_{c'}
\ee
with $d_{a_1\dots a_k}$ defined by \eqref{inv}.
The invariants depending only on the angular momentum variables
agree with ${\cal L}_{2k}$ as defined in~\eqref{Lk}.
They are complemented to a full set of integrals
by the following simple family of invariants,
\be
\label{An}
\mathcal{A}_k=\text{tr}\,\mathbf{A}^k = \sum_{a_1,\dots,a_k} d_{a_1\dots a_k}A_{a_1}\ldots A_{a_k}.
\ee

Let us write down explicit expressions for the first two integrals from this family. Using the representation
\eqref{Aa-2}, we immediately get
 \be
 \label{A1}
{\cal A}_1=\left(\frac{2x_0}{r_0}\mathcal{H}_\gamma+\frac{x_0}{r_0^3}(\bm{x}\cdot\bm{p})^2-\frac{2\gamma x}{r_0^2}+\frac{\gamma}{x}\right) {\cal X}
-\frac{x_0}{r_0}(\bm{x}\cdot\bm{p}) {\cal P},
\ee
where ${\cal X}$ and ${\cal P}$ are defined in~\eqref{calXP}.
As a direct consequence of~\eqref{Asq},
the second integral  is expressed via the Hamiltonian \eqref{HrCal} and the angular
Calogero Hamiltonian,
\be
{\cal A}_2=\left({\cal H}_\gamma-\frac{{\cal L}_2}{4r_0^2}\right){\cal L}_2+\gamma^2.
\ee
Thus, using the method of matrix-model reduction, we constructed a
complete set of constants of motion for the Calogero-oscillator
and Calogero-Coulomb models on the $N$-dimensional sphere.

\section{Generalizations}

\subsection{Calogero-Coulomb-Stark on the sphere}
\noindent
Consider an integrable  generalization of the Coulomb-Stark system  to the sphere \cite{anosc},
\be
H_{\gamma,F}=H_{\gamma}+\frac{x_0}{r_0}(\bm{F}\cdot\bm{x}),
\label{Stark}
\ee
where $H_\gamma$ is given by~\eqref{Hr} and
$\bm{F}$ is an analog of  the constant electric  field in the planar limit.
This system lacks  the superintegrability but still remains integrable.
Its constants of motion are given by the  angular-momentum components
 orthogonal to the electric field direction,
\be
\label{Lbot}
L^{\bot}_{ij}= L_{ij}+n_in_kL_{jk}-n_jn_kL_{ik},
\ee
and a modified longitudinal component of the Runge-Lenz vector on the sphere \eqref{Aa} given by
\cite{anosc}
 \be
 \label{AStark}
 A=\bm{n}\cdot\bm{A}- \frac{F}{2}\left(\bm{x}^2-(\bm{n}\cdot\bm{x})^2\right),
\ee
where we introduced the field strength $F=|\bm{F}|$ and the field direction $\bm{n}=\bm{F}/F$.

The matrix-model reduction described in the previous section can be applied  to the extension of the system \eqref{Stark}
by  a Calogero potential.
The electric field is now characterized by a matrix proportional to the identity, to ensure
the SU$(N)$ invariance of the Stark term under the adjoint action \eqref{SUn}:
\be
\mathbf{F}=\frac{F}{\sqrt{N}}\mathbf{1}=\frac{F}{\sqrt{N}} \sum_{k=1}^N \mathbf{T}_k,
\qquad
\text{tr}\,\mathbf{F}\mathbf{X}= \frac{F}{\sqrt{N}} \text{tr}\,\mathbf{X}.
\ee
Therefore, the matrix-model analog of the system on the sphere \eqref{Stark}  takes the form
\be
\label{Hf-mat}
\mathcal{H}^{\rm mat}_{\gamma,F}
= \mathcal{H}^{\rm mat}_\gamma+\frac{F}{\sqrt{N}}\frac{x_0}{r_0}\;\text{tr}\;\mathbf{X}
=\mathcal{H}^{\rm mat}_\gamma +\frac{F}{\sqrt{N}} \frac{x_0}{r_0}\sum_{k=1}^N X_k,
\ee
with $ \mathcal{H}^{\rm mat}_\gamma$ given by \eqref{Hr-mat}.
In other words, the field matrix is tangent to the U$(1)$ center of the unitary group, and
the unit vector along that direction is given by the invariant vector \eqref{inv12} as
$n_a=d_a/\sqrt{N}$. So, the conserved transversal components of the angular momentum
\eqref{Lbot} acquire the form
\be
\label{Lbot-ab}
L^{\bot}_{ab}= L_{ab}+\frac{1}{N}\sum_{k=1}^N (d_a L_{bk}-d_b L_{ak}),
\ee
where the first equation in~\eqref{inv12} has been used.

The modified component of the Runge-Lenz vector tangent to the U$(1)$ center is given by
\be
\label{Apar-mat}
A=\frac{1}{\sqrt{N}}\text{tr}\;\mathbf{A}  -\frac{F}{2}\Big( \text{tr}\;\mathbf{X}^2 -\frac{1}{N} \big( \text{tr}\;\mathbf{X} \big)^2\Big).
\ee

Reducing the matrix model \eqref{Hf-mat} by the SU$(N)$ group action, we  arrive at
the  Calogero-Coulomb-Stark Hamiltonian on the sphere:
\be
\label{Hf}
\mathcal{H}_{\gamma,F}=\mathcal{H}_\gamma+ \frac{F}{\sqrt{N}}\frac{x_0}{r_0 }{\cal X}
\ee
with ${\cal X}$ given in~\eqref{calXP}.

Projecting the matrix-model integrals \eqref{Lbot-ab} and \eqref{Apar-mat} to the
SU$(N)$ invariant orbits in the usual way, we arrive at the following constants of motion
for the above Hamiltonian,
\begin{align}
\mathcal{L}^\bot_{2k}
&=\sum_{a_1,b_1\dots,a_{2k},b_{2k}} d_{a_1\dots a_{2k}}d_{b_1\dots b_{2k}}L^\bot_{a_1b_1}\ldots L^\bot_{a_{2k}b_{2k}},
\\
\mathcal{A}&=\frac{1}{\sqrt{N}}{\cal A}_1  -\frac{F}{2}\Big( \bm{x}^2 -\frac{{\cal X}^2}{N}\Big),
\end{align}
where the definition \eqref{A1} is taken into account.
The above constants of motion ensure the integrability of the system.

In the $r_0\to\infty$ limit the Hamiltonian \eqref{Hf} describes to the Calogero-Coulomb-Stark problem in flat space,
which has been introduced and studied in~\cite{Calogero-Stark} using the Dunkl-operator approach.

\subsection{Spin extensions}
\noindent
The  aforementioned systems on the sphere can be endowed  with an additional classical spin
while retaining their integrability or superintegrability.
Such extensions have been studied in the flat-space limit
for the Calogero model  using the Lax-pair technique, by introducing internal degrees of freedom $l_{ij}$
into the inverse-square potential of the system~\eqref{Hx-red}~\cite{woj-euler}.
In particular, the Calogero-oscillator Hamiltonian on the sphere \eqref{Hx-mat}
with classical spins has the following form:
\be
\label{Hx-spin}
\mathcal{H}^\text{spin}_\omega=\frac{\bm{p}^2}{2}-\frac{(\bm{x}\cdot\bm{p})^2}{2r_{0}^2}
+\sum_{i<j}\frac{l_{ij}^2}{(x_i-x_j)^2}+\frac{\omega^2r_{0}^2}{2}\frac{x^2}{x_0^2}.
\ee
The spin dynamic variables $l_{ij}$ obey $so(N)$ angular momentum Poisson bracket relations \eqref{poisLij},
but  they are in involution with the variables $p_i$ and $x_i$ describing the motion on the sphere.
They can be recast into a spin degree of freedom $\bm{S}$ related to each coordinate:
$l_{ij}=\bm{S}_i\cdot\bm{S}_j$ \cite{poly-min}.

Actually, the classical spin $l_{ij}$ can be obtained also from the matrix-model reduction procedure
\cite{calogero-review,poly97,poly-min}.
Remember that the Noether generators \eqref{J} form a traceless Hermitian matrix, and so far we took the minimal
gauge~\eqref{Jijcal} for them. Let us instead apply a spin gauge by imposing  a weaker condition:  $J_{ij}=\imath l_{ij}$.
Using the antisymmetry of the $l_{ij}$, the conventional Lax matrix~\eqref{XPred} gets replaced by
\be
\label{XPredspin}
P_{ij}=
p_i\delta_{ij}-(1-\delta_{ij})\frac{l_{ij}}{x_i-x_j} .
\ee

The construction scheme for the integrals of motion described in Section~3 can be extended to the spin case too.
The integrals are given by the SU$(N)$ invariant polynomials built from the spin matrix $l_{ij}$ and the reduced matrices
$X_{ij}$ and $P_{ij}$. In the flat-space limit the algebra of integrals of the Calogero system with and without
oscillator term has been studied in this context in~\cite{avan}.

\section{Concluding remarks}
\noindent
We have defined the classical rational Calogero model with an oscillator or Coulomb potential
on the sphere or hyperboloid. This may be viewed as the oscillator or Coulomb system
on the sphere or hyperboloid, as introduced, respectively, by Higgs and Schr\"odinger~\cite{higgs,schroedinger},
amended by a Calogero interaction term.
This system looks similar to but is distinct from the angular part of the Calogero model
in the ambient $(N{+}1)$-dimensional flat space.
Both systems however share features such as superintegrability, a Lax pair and matrix-model descriptions.

We have focused on the constants of motion for the Calogero-oscillator and Calogero-Coulomb systems on the sphere.
They were obtained from the kinematical and hidden dynamical symmetries of the related Hermitian matrix models
by a Hamiltonian reduction.
We have expressed the constants of motion of the reduced systems  as  SU$(N)$ invariant polynomials
depending on the well known integrals of the original matrix model.
We also have studied the spherical generalization of the Calogero-Coulomb system in an external electric field,
i.e.\ with a Stark term,
and briefly discussed the effect of additional spin degrees of freedom.

In a forthcoming  paper we will extend the construction carried out here to the quantum case~\cite{quantum}.
It will be interesting to consider in this context the spherical Calogero models
associated with general Coxeter root systems.

\acknowledgments
\noindent
T.H.\ and A.N.\ were partially supported by the Armenian
State Committee of Science Grants No.~15RF-039 and No.~15T-1C367 and by Grant No.~mathph-4220 of the
Armenian National Science and Education Fund based in New York (ANSEF). The work of T.H.\ and A.N.\
was done within the ICTP programs NET68 and OEA-AC-100 and within the program of Regional Training Networks 
on Theoretical Physics by VolkswagenStiftung contract nr.~86~260.
The work of F.C.\ is supported by the Alexander von Humboldt Foundation under grant CHL 1153844 STP.


\begin{thebibliography}{99}

\bibitem{calogero-review}
A.P.~Polychronakos,
\emph{Physics and mathematics of Calogero particles},
J. Phys. A  {\bf 39} (2006) 12793,
\href{http://arxiv.org/abs/hep-th/0607033}{hep-th/0607033}.

\bibitem{calogero0}
F.~Calogero,
\emph{Solution of a three-body problem in one-dimension},
\href{http://dx.doi.org/10.1063/1.1664820}{J. Math. Phys. {\bf 10} (1969) 2191};
\emph{Solution of the one-dimensional N-body problems with quadratic and/or inversely quadratic
pair potentials},
\href{http://dx.doi.org/10.1063/1.1665604}{{\sl ibid.} {\bf 12} (1971) 419}.

\bibitem{moser}
J. Moser,
\emph{Three integrable Hamiltonian systems connected with isospectral deformations},
\href{http://dx.doi.org/10.1016/0001-8708(75)90151-6}{Adv. Math. {\bf 16} (1975) 197}.

\bibitem{calogero-root}
M.A.~Olshanetsky and  A.M.~Perelomov,
 \emph{Classical integrable finite dimensional systems related to Lie algebras},
\href{http://dx.doi.org/10.1016/0370-1573(81)90023-5}
{Phys. Rept.  {\bf 71} (1981) 313};
\emph{Quantum integrable systems related to Lie algebras},
\href{http://dx.doi.org/10.1016/0370-1573(83)90018-2 }{{ibid.}  {\bf 94} (1983) 313}.

\bibitem{woj83}
 S.~Wojciechowski,
\emph{Superintegrability of the Calogero-Moser system},
\href{http://dx.doi.org/10.1016/0375-9601(83)90018-X}{Phys. Lett. A {\bf 95} (1983) 279}.

\bibitem{gonera98}
 C. Gonera,
 \emph{On the superintegrability of Calogero-Moser-Sutherland model},
\href{http://dx.doi.org/10.1088/0305-4470/31/19/012}{J. Phys. A {\bf 31} (1998) 4465}.

\bibitem{sutherland}
B. Sutherland,
\emph{Exact results for a quantum many-body problem in one dimension},
\href{http://dx.doi.org/10.1103/PhysRevA.4.2019}{Phys. Rev. A {\bf 4} (1971) 2019};
\emph{Exact results for a quantum many body problem in one dimension. II},
\href{http://dx.doi.org/10.1103/PhysRevA.5.1372}{ibid. {\bf 5} (1972) 1372}.

\bibitem{poly97}
A.~Polychronakos,
\emph{Multidimensional Calogero systems from matrix models},
Phys.  Lett. B {\bf 408} (1997) 117,
\href{http://arxiv.org/abs/hep-th/9705047}{hep-th/9705047}.

\bibitem{CalCoul}
  T.~Hakobyan, O.~Lechtenfeld, and A.~Nersessian,
\emph{Superintegrability of generalized Calogero models with oscillator or Coulomb potential},
   Phys. Rev. D \textbf{90}  (2014) 101701(R),
 \href{http://arxiv.org/abs/1409.8288}{arXiv:1409.8288}.

\bibitem{Runge}
T.~Hakobyan and A.~Nersessian,
\emph{Runge-Lenz vector in Calogero-Coulomb problem},
 Phys. Rev. A \textbf{92} (2015) 022111,
\href{http://arxiv.org/abs/1504.00760}{arXiv:1504.00760}.

\bibitem{adler}
M. Adler,
\emph{Some finite dimensional integrable systems and their scattering behavior},
 \href{http://dx.doi.org/10.1007/BF01614548}{Comm. Math. Phys. {\bf 55} (1977) 195}

C. Gonera and P. Kosinski,
 \emph{Calogero model and $sl(2,R)$ algebra},
 Acta Phys. Polon. B {\bf 30} (1999) 907,
 \href{http://arxiv.org/abs/hep-th/9810255}{hep-th/9810255}.

\bibitem{Calogero-Stark}
T. Hakobyan and A. Nersessian,
\emph{Integrability and separation of variables in Calogero-Coulomb-Stark and two-center
Calogero-Coulomb systems}, Phys. Rev.  D \textbf{93} (2016) 045025,
\href{http://arxiv.org/abs/1509.01077}{arXiv:1509.01077}.

\bibitem{schroedinger}
E. Schr\"odinger,
\emph{A method of determining quantum-mechanical eigenvalues and eigenfunctions},\\
\href{http://www.jstor.org/stable/20490744}{Proc. Roy. Ir. Acad. A {\bf 46} (1940) 9}.

\bibitem{higgs}
P. W. Higgs,
\emph{Dynamical symmetries in a spherical geometry. I},
\href{http://dx.doi.org/10.1088/0305-4470/12/3/006}{J. Phys. A {\bf 12} (1979) 309}.

\bibitem{gns}
  A.~Galajinsky, A.~Nersessian and A.~Saghatelian,
 \emph{Superintegrable models related to near horizon extremal Myers-Perry black hole in arbitrary dimension},
  JHEP {\bf 1306}, 002 (2013),
\href{http://arxiv.org/abs/1303.4901}{arXiv:1303.4901}.

\bibitem{sphCal}
T. Hakobyan, D. Karakhanyan, and O. Lechtenfeld,
\emph{The structure of invariants in conformal mechanics},
 Nucl. Phys. B {\bf 886} (2014) 399,
\href{http://arxiv.org/abs/1402.2288}{arXiv:1402.2288};
T.~Hakobyan, S.~Krivonos, O.~Lechtenfeld and A.~Nersessian,
\emph{Hidden symmetries of integrable conformal mechanical systems},
Phys. Lett.  A {\bf 374} (2010) 801,
\href{http://arxiv.org/abs/0908.3290}{arXiv:0908.3290};
T.~Hakobyan, A.~Nersessian and V.~Yeghikyan,
J.  Phys. A  {\bf 42}  (2009) 205206,

\bibitem{sph-mat}
T.~Hakobyan, O.~Lechtenfeld and A.~Nersessian,
\emph{The spherical sector of the Calogero model as a reduced matrix model},
Nucl. Phys. B  {\bf 858} (2012) 250,
\href{http://arxiv.org/abs/1110.5352}{arXiv:1110.5352}.

\bibitem{flp}
M.~Feigin, O.~ Lechtenfeld, and A.~Polychronakos,
\emph{The quantum angular Calogero-Moser model},
JHEP {\bf 1307} (2013) 162,
\href{http://arxiv.org/abs/1305.5841}{arXiv:1305.5841};
 F.~Correa and O.~Lechtenfeld, \emph{The tetrahexahedric angular Calogero model},
  JHEP {\bf 1510} (2015) 191,
  \href{http://arxiv.org/abs/1508.04925}{arXiv:1508.04925}.

\bibitem{Feigin}
  M.~Feigin and T.~Hakobyan,
\emph{On Dunkl angular momenta algebra},
  JHEP {\bf 1511} (2015) 107,
  \href{http://arxiv.org/abs/1409.2480}{arXiv:1409.2480}.

\bibitem{fradkin}
D.M. Fradkin,
\emph{Existence of the dynamic symmetries $O_4$ and $SU_3$ for all classical central potential problems},
\href{http://dx.doi.org/doi:10.1143/PTP.37.798}{Theor. Phys. {\bf 37} (1967) 798};
\emph{Three-dimensional isotropic harmonic oscillator and $SU_3$},
\href{http://dx.doi.org/doi:10.1119/1.1971373}{Am. J. Phys. {\bf 33} (1965) 207}.

\bibitem{kazhdan}
D.~Kazhdan, B.~Kostant and S.~Sternberg,
\emph{Hamiltonian group actions and dynamical systems of Calogero type},
\href{http://dx.doi.org/10.1002/cpa.3160310405}{Comm. Pure Appl. Math. {\bf 31} (1978) 481}.

\bibitem{anosc}
  A.~Nersessian and V.~Yeghikyan,
  \emph{Anisotropic inharmonic Higgs oscillator and related (MICZ-) Kepler-like systems},
  J. Phys. A {\bf 41} (2008) 155203,
 \href{http://arxiv.org/abs/0710.5001} {arXiv:0710.5001};
\emph{Anisotropic Higgs oscillator},
 Proc. of VII  International Workshop on Supersymmetries and Quantum Symmetries (SQS'07),
30.07-4.08.2007,  Dubna,
  \href{http://arxiv.org/abs/arXiv:0711.1033}{arXiv:0711.1033};
  S.~Bellucci and V.~Yeghikyan,
  \emph{The Coulomb problem on a 3-sphere and Heun polynomials},
  J. Math. Phys.  {\bf 54}, 082103 (2013)
  \href{http://arxiv.org/abs/1302.0798}{arXiv:1302.0798}.

\bibitem{woj-euler}
S. Wojciechowski,
\emph{An integrable marriage of the Euler equations with the Calogero-Moser system},\\
\href{http://dx.doi.org/doi:10.1016/0375-9601(85)90432-3}{Phys. Lett. A {\bf 111} (1985) 101};
J. Gibbons and T. Hermsen,
\emph{A generalisation of the Calogero-Moser system},\\
\href{http://dx.doi.org/doi:10.1016/0167-2789(84)90015-0}{Physica D {\bf 11} (1984) 337}.

\bibitem{poly-min}
J.A. Minahan and A.P. Polychronakos,
\emph{Interacting fermion systems from two dimensional QCD},
Phys. Lett. B {\bf 326} (1994) 288,
\href{http://arxiv.org/abs/hep-th/9309044}{hep-th/9309044}.

\bibitem{avan}
J. Avan and E. Billey,
\emph{Observable algebras for the rational and trigonometric Euler-Calogero-Moser models},\\
Phys. Lett. A {\bf 198} (1995) 183,
\href{http://arxiv.org/abs/hep-th/9404040}{hep-th/9404040};
J. Avan,
\emph{Integrable extensions of the rational and trigonometric $A_N$ Calogero-Moser potentials},
Phys. Lett. A {\bf 185} (1994) 293,
\href{http://arxiv.org/abs/hep-th/9306112}{hep-th/9306112}.

\bibitem{quantum}
F. Correa, T. Hakobyan, O. Lechtenfeld, and A. Nersessian,
\emph{Spherical Calogero model with oscillator/Coulomb potential: quantum case},
work in preparation.

\end{thebibliography}
\end{document}